\documentclass[review]{elsarticle}

\usepackage{lineno,hyperref}
\modulolinenumbers[1]

\journal{Astronomy and Computing}

\usepackage[margin=0.65in]{geometry}

\begin{document}

\begin{frontmatter}

\title{\texttt{easyFermi}: a graphical interface for performing \textit{Fermi}-LAT data analyses}

\author{Raniere de Menezes$^{1,2}$\corref{mycorrespondingauthor}}
\address{$^{1}$Universidade de S\~ao Paulo, Departamento de Astronomia, Rua do Mat\~ao, 1226, S\~ao Paulo, SP 05508-090, Brazil\\
$^{2}$Lehrstuhl f\"ur Astronomie, Universit\"at W\"urzburg, Emil-Fischer-Strasse 31, 97074 W\"urzburg, Germany\\
\textbf{Submitted to Astronomy and Computing on 07/March/2022, accepted on 22/June/2022}\\}
\cortext[mycorrespondingauthor]{E-mail: raniere.m.menezes@gmail.com}




\begin{abstract}
Since its launch in 2008, the \textit{Fermi} Large Area Telescope (LAT) allowed us to peek into the extremely energetic side of the Universe with unprecedented sensitivity and resolution. The tools available for analyzing \textit{Fermi}-LAT data are the \texttt{Fermitools} and \texttt{Fermipy}, both of which can be scripted in Python and run via command lines in a terminal or in web-based interactive computing platforms. In this work, we are providing the community with \texttt{easyFermi}, an open-source user-friendly graphical interface for performing basic to intermediate analyses of \textit{Fermi}-LAT data in the framework of \texttt{Fermipy}. With \texttt{easyFermi}, the user can quickly measure the $\gamma$-ray flux and photon index, build spectral energy distributions, light curves, test statistic maps, test for extended emission and even relocalize the coordinates of $\gamma$-ray sources. The tutorials for \texttt{easyFermi} are available on YouTube and GitHub, allowing the user to learn how to use \textit{Fermi}-LAT data in about 10 min.
\end{abstract}


\end{frontmatter}


\setlength{\columnsep}{30pt}
\twocolumn

\section{Introduction}

The Large Area Telescope (LAT) onboard the \textit{Fermi Gamma-ray Space Telescope} (\textit{Fermi}) is a pair-conversion telescope with imaging and spectroscopic capabilities, a large field-of-view of $\sim 2$ sr, and sensitive to $\gamma$-rays in the energy range from $\sim 20$ MeV to $\sim 1$ TeV\footnote{\url{https://www.slac.stanford.edu/exp/glast/groups/canda/lat_Performance.htm}} \citep{atwood2009large}. The \textit{Fermi}-LAT performs an all-sky survey every $\sim 3$ hours and, since its launch in 2008, has collected more than 13 years of data.

Once the events detected by the \textit{Fermi}-LAT are classified, the collected $\gamma$-ray data are released to the public on the Fermi Science Support Center (FSSC) data server\footnote{\url{https://fermi.gsfc.nasa.gov/cgi-bin/ssc/LAT/LATDataQuery.cgi}}. The users of these data can rely on \texttt{Fermitools}\footnote{\url{https://fermi.gsfc.nasa.gov/ssc/data/analysis/software/}} and \texttt{Fermipy}\footnote{\url{https://fermipy.readthedocs.io/en/latest/}} \citep{wood2017fermipy} to perform their analyses, both tools relying on several online tutorials\footnote{\url{https://fermi.gsfc.nasa.gov/ssc/data/analysis/scitools/}, \url{https://fermipy.readthedocs.io/en/latest/quickstart.html}}. \texttt{Fermipy}, in particular, is a high-level \texttt{Python} package that facilitates the analysis of \textit{Fermi}-LAT data in the framework of the \texttt{Fermitools}.

In this work we present \texttt{easyFermi}, an open-source graphical interface suited to perform \textit{Fermi}-LAT data analyses of point-like and extended $\gamma$-ray sources. \texttt{easyFermi} facilitates the experience and can save a substantial amount of time for those doing $\gamma$-ray astronomy, especially to scientists just starting in the field of high-energy astrophysics. Here we provide an introduction to what is possible to do with \texttt{easyFermi} and how to use its graphical interface. Further tutorials can be found online on GitHub\footnote{\url{https://github.com/ranieremenezes/easyFermi}} and YouTube\footnote{\url{https://www.youtube.com/channel/UCeLCfEoWasUKky6CPNN_opQ}}. This paper is organized as follows. We describe the installation and setup processes of \texttt{easyFermi} in \S \ref{sec:installation} and detail what is happening behind the scenes, especially the dependency of \texttt{easyFermi} on \texttt{Fermipy}, in \S \ref{sec:behind}. In \S \ref{sec:results} we show some of the main data products of \texttt{easyFermi} and, finally, we conclude in \S \ref{sec:conclusions}.

\section{Installation and setup}
\label{sec:installation}

The current (and upcoming) release of \texttt{easyFermi} (i.e. V1.0.7) is available in the Python Package Index (PyPI) server\footnote{\url{https://pypi.org/project/easyFermi/}} and GitHub, together with detailed instructions for installation and usage. The installation of \texttt{easyFermi} (V1.0.7) requires an existing installation of the \texttt{Fermitools} V2.0.8 and \texttt{Fermipy} V1.0.1 (Python 3 version) and works on Linux and Mac operational systems. The graphical interface will be maintained such that its installation will be compatible with the upcoming releases of the \texttt{Fermitools} and \texttt{Fermipy}.

In summary, once the \texttt{Fermitools} and \texttt{Fermipy} are installed (see the online tutorial in the \texttt{easyFermi} GitHub webpage), one can use the terminal to open the \texttt{fermi} environment with \texttt{conda} by typing:

\texttt{\$ conda activate fermi}

And then simply type:

\texttt{\$ pip install easyFermi}

To test if the installation is properly working, the user can type:

\texttt{\$ python}

\texttt{>> import easyFermi}

\begin{figure*}
    \centering
    \includegraphics[width=\linewidth]{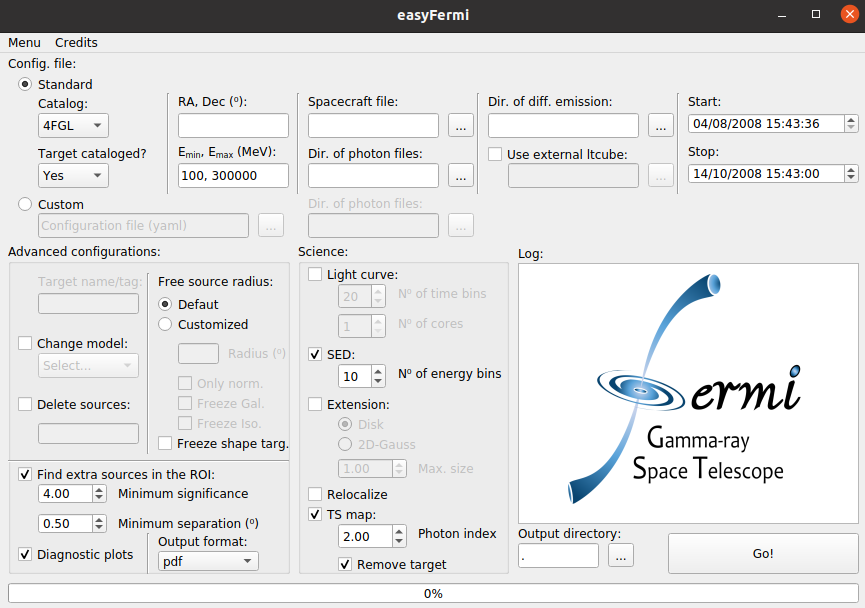}
    \caption{The main window of \texttt{easyFermi}.}
    \label{fig:easyFermi}
\end{figure*}

{\noindent The graphical interface of \texttt{easyFermi} should appear at this point (see Figure \ref{fig:easyFermi}) and the user can enter the desired configurations.}

To exemplify the usage of the software, we will use the blazar PG 1553+113 as a test target. The first step is to download the spacecraft and photon data files from the FSSC server\footnote{\url{https://fermi.gsfc.nasa.gov/cgi-bin/ssc/LAT/LATDataQuery.cgi}}, as well as the Galactic (\texttt{gll\_iem\_v07}) and isotropic (iso\_P8R3\_SOURCE\_V3\_v1) diffuse emission models\footnote{\url{https://fermi.gsfc.nasa.gov/ssc/data/access/lat/BackgroundModels.html}}. The parameters that we adopt to download the \textit{Fermi}-LAT data for PG 1553+113 are:

\begin{itemize}
    \item Coordinates: 238.92935, 11.19010 ($^{\circ}$)
    \item Search radius: 10 ($^{\circ}$)
    \item Observation dates: 2008-08-04 15:43:36, 2009-08-04 00:00:00 (Gregorian)
    \item Energy range: 1000, 500000  (MeV)
\end{itemize}

For a standard point-source analysis of PG 1553+113, the user only needs to feed the empty boxes of \texttt{easyFermi} with the coordinates (i.e. 238.92935, 11.19010), desired energy range (1000, 500000), the directory where the spacecraft, photon and background files were downloaded, and the adopted time interval (from 2008-08-04 15:43:36 to 2009-08-04 00:00:00). For the users with some experience on \textit{Fermi}-LAT data analysis, there are a set of advanced configurations that can also be controlled, but all of them are optional. Since PG 1553+113 is listed in 4FGL, the configuration step is finished. For targets not listed in 4FGL (or 3FGL), the user has to change the ``Target cataloged'' entry to ``No'' and insert a nickname for the target in the box ``Target name/tag''.

The user can then simply choose what are the desired outputs by checking the Light curve, Spectral Energy Distribution (SED), Extension, Re-localize and test statistic (TS) map boxes. The results will then be saved as ``.txt'', ``.npy'', ``.fits'', and ``.pdf'' (or ``.png'') files in the selected output directory. If no output directory is chosen, \texttt{easyFermi} will save all files in the current working directory. The analysis will start after pressing ``Go!'', however, the user can still make modifications on the fly, like activating or deactivating one of the boxes in the \texttt{Science} panel. For more complex analysis, we refer the user to the online tutorials.

\section{Behind the scenes}
\label{sec:behind}

The graphical interface and data analysis provided by \texttt{easyFermi} strongly depend on the \texttt{Python} packages \texttt{PyQt5} and \texttt{Fermipy}, although requiring minimum maintenance. Once the user sets the desired configurations and start the process, \texttt{easyFermi} automatically organizes all these information with \texttt{PyQt5} and communicates them to \texttt{Fermipy}, thereafter starting a binned likelihood analysis. This process generates several data subproducts, as a list of selected events, a counts cube, an exposure map, a livetime cube, and a source map\footnote{More details on these data subproducts here: \url{https://fermi.gsfc.nasa.gov/ssc/data/analysis/scitools/binned_likelihood_tutorial.html}}.  

Before fitting the model to the data in the region of interest (ROI), we first optimize the ROI model with the \texttt{Fermipy} function \texttt{optimize()} to ensure that all parameters are close to their global likelihood maxima, and then look for uncataloged sources with the function \texttt{find\_sources()} (this one can be disabled by the user). The model is then fitted to the data using the minimizer \texttt{MINUIT} and the main results for the target are saved in the file ``Target\_results.txt''. For the full set of results including all sources in the ROI, the user can access the files ``Results.fits'' or ``Results.npy''.

\subsection{Current spectral models}

In \texttt{easyFermi} we use the spectral models from the \textit{Fermi}-LAT third and fourth source catalogs \citep[3FGL and 4FGL, respectively;][]{acero2015_3FGL,abdollahi2020_4FGL}, namely i) the power-law model, defined by $$ \frac{dN}{dE} = N_0 \left( \frac{E}{E_0} \right)^{\gamma}, $$ where $N_0$ is the normalization (in units of cm$^{-2}$ s$^{-1}$ MeV$^{-1}$), $E$ is the photon energy, $E_0$ is the pivot energy, and $\gamma$ is the photon index; ii) the log-parabolic model \citep{massaro2004log} defined by $$ \frac{dN}{dE} = N_0 \left( \frac{E}{E_0} \right)^{-\alpha -\beta\log(E/E_0)},$$ where $\alpha$ and $\beta$ are indexes describing the hardness and curvature of the spectrum; and iii) the power-law with a super exponential cutoff $$ \frac{dN}{dE} = N_0 \left( \frac{E}{E_0} \right)^{\gamma}\exp (-aE^b),$$ where $a$ and $b$ are the exponential factor and index describing the shape of the spectral cutoff. If the target is listed in one of the \textit{Fermi}-LAT catalogs, \texttt{easyFermi} automatically uses the cataloged spectral model in the analysis, otherwise a new source with a power-law spectrum is added to the ROI model. The user can always change the spectral model for the target under the checkbox ``Change model''. Depending on the feedback by the users of \texttt{easyFermi}, we can add more spectral models in the next software releases.

\subsection{Automatic configuration}

For those using the \texttt{Standard} mode of \texttt{easyFermi}, the following set of configurations apply. 
\begin{itemize}
    \item The ROI is defined as an $L \times L$ square with size depending on the adopted starting energy ($E_{min}$), i.e. $L = 15^{\circ}, 12^{\circ}$ or $10^{\circ}$ for $E_{min} < 500$ MeV, 500 MeV $\leq E_{min} < 1000$ MeV and $E_{min} \geq 1000$ MeV, respectively.
    \item For the same energy ranges defined in the previous item, we set the maximum zenith angle to $z_{max} = 90^{\circ}, 100^{\circ}$ or $105^{\circ}$.
    \item The classification and point spread function type for each event are filtered with \texttt{evclass} $= 128$ and \texttt{evtype} $= 3$, while the adopted instrument response function is \texttt{P8R3\_SOURCE\_V3} and the dataset is divided in 8 bins per energy decade.
    \item The radius from the ROI center in which the parameters of the sources are allowed to vary during the fit is defined as $R_{free} = L/2$, where $L$ is the size of the ROI.
\end{itemize}

The experienced user can change all of these configurations by passing a customized configuration file to \texttt{easyFermi} under the selection of the ``custom'' button and by modifying the entries in the ``Advanced configurations'' box (see Figure \ref{fig:easyFermi}).

\subsection{Recovering the state of the analysis}

A very useful characteristic of \texttt{easyFermi} is that it allows the user to recover the latest state of the analysis, such that the user can quit \texttt{easyFermi} and continue the exact same analysis some time later. Once the analysis is done, the state of the graphical interface is automatically saved in the file ``GUI\_status.npy'' in the output directory, and can easily be recovered by clicking on ``Load GUI state'' in the ``Menu'' button of the toolbar.

\section{Main data products}
\label{sec:results}

The main results from \texttt{easyFermi} are the measurements of the flux and spectral shape for all sources in the ROI, which are saved in the files ``Target\_results.txt'', ``Results.npy'', and ``Results.fits''. Furthermore, the user can relocalize the target (the ROI is then updated with the new location of the target), compute a light curve, a $\gamma$-ray spectrum, look for extended emission and compute a TS map.

\begin{figure}
    \centering
    \includegraphics[width=\linewidth]{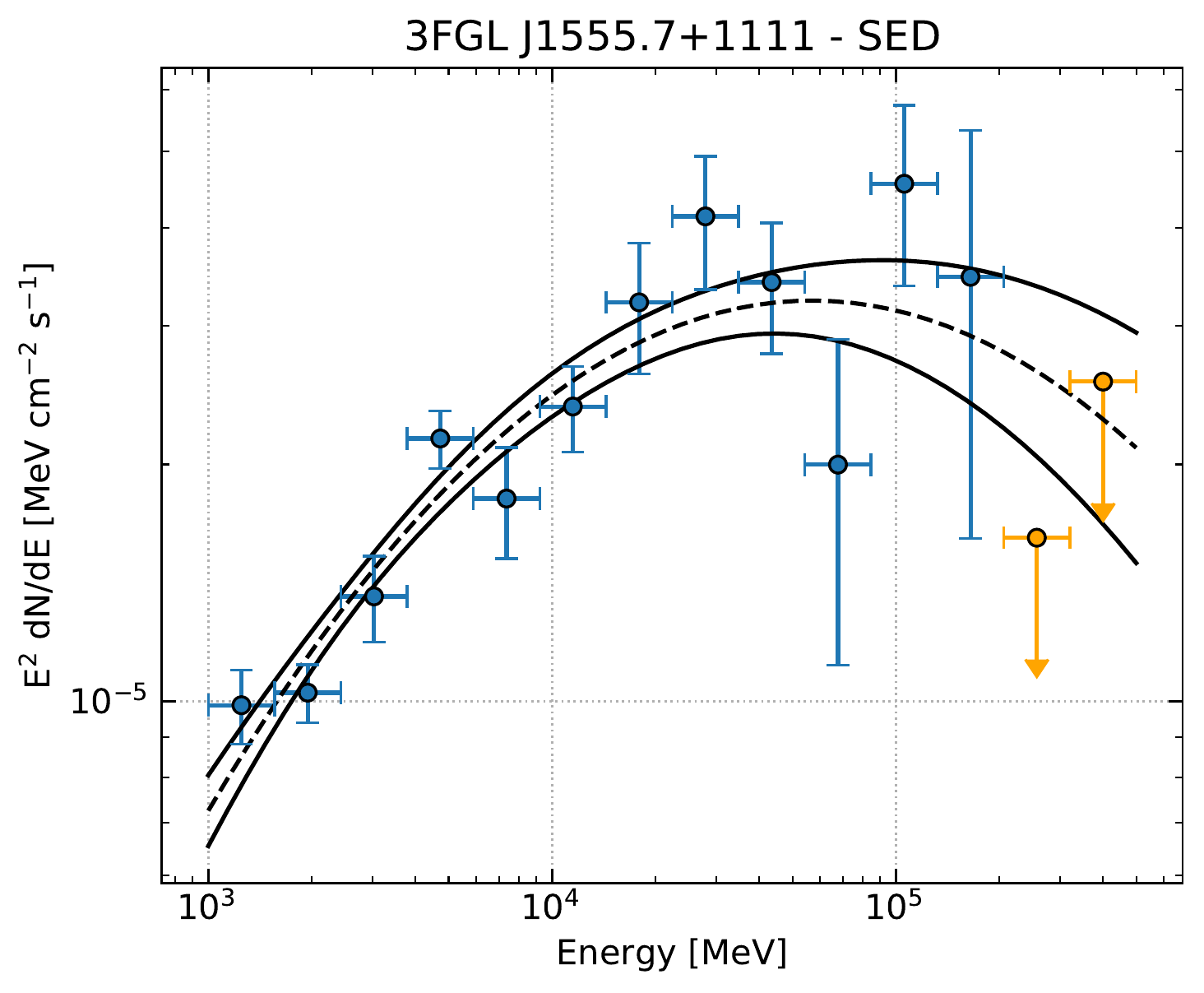}
    \caption{The $\gamma$-ray spectrum for PG 1553+113 fitted with a log-parabolic model.}
    \label{fig:SED}
\end{figure}

\begin{figure*}
    \centering
    \includegraphics[width=\linewidth]{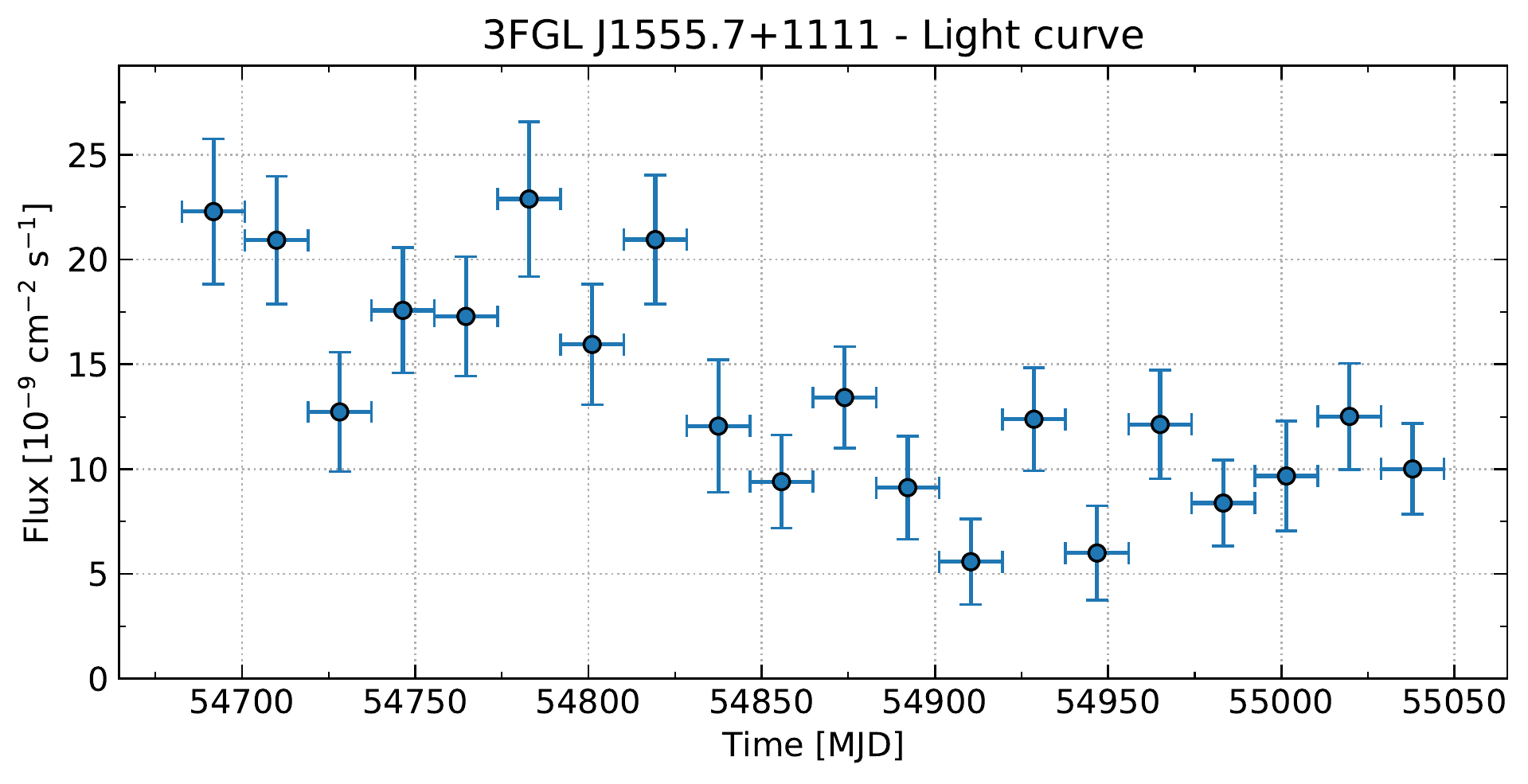}
    \caption{Photon flux light curve for PG 1553+113.}
    \label{fig:LC}
\end{figure*}

\begin{figure*}
    \centering
    \includegraphics[width=\linewidth]{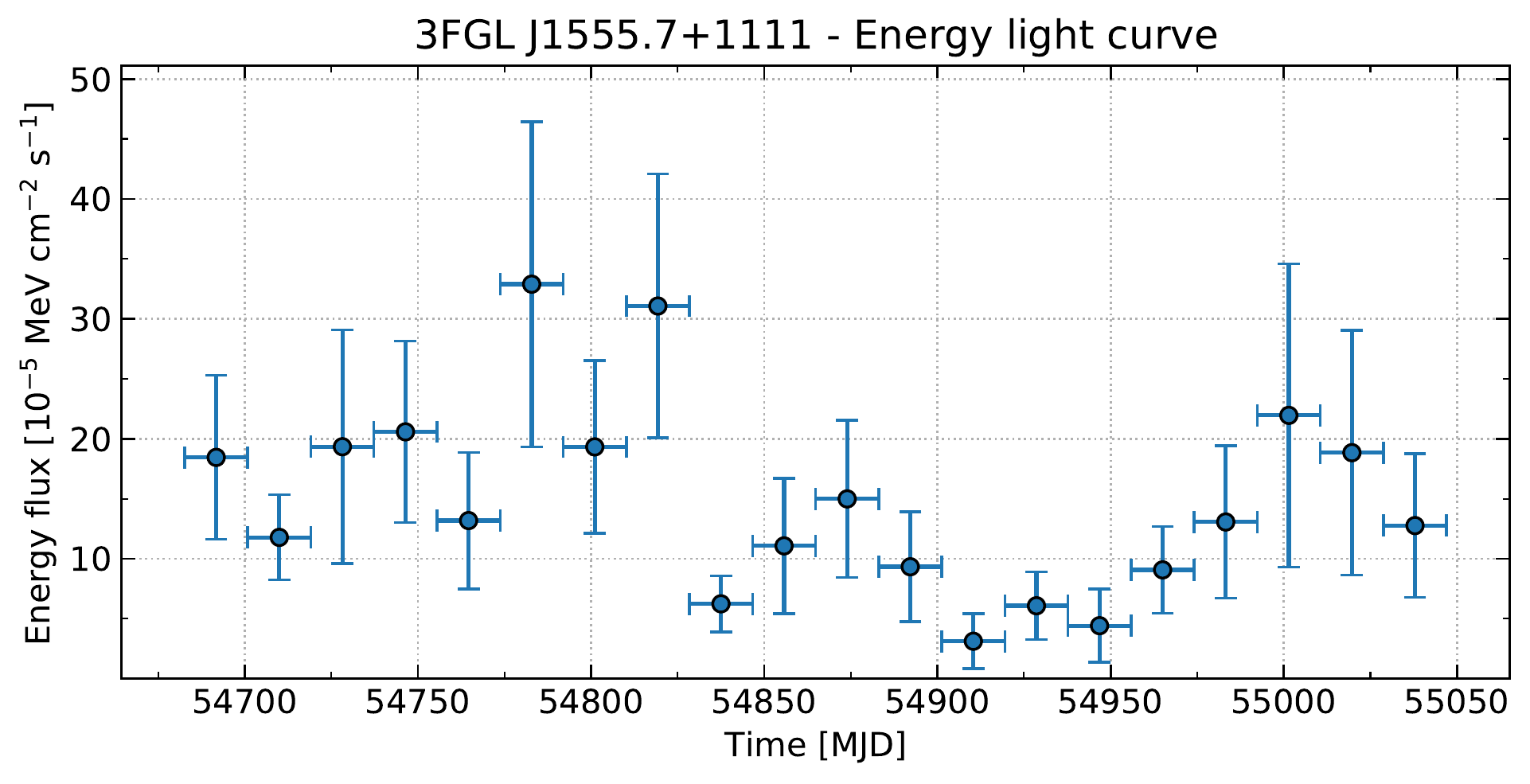}
    \caption{Energy flux light curve for PG 1553+113.}
    \label{fig:eLC}
\end{figure*}

\begin{figure}
    \centering
    \includegraphics[width=\linewidth]{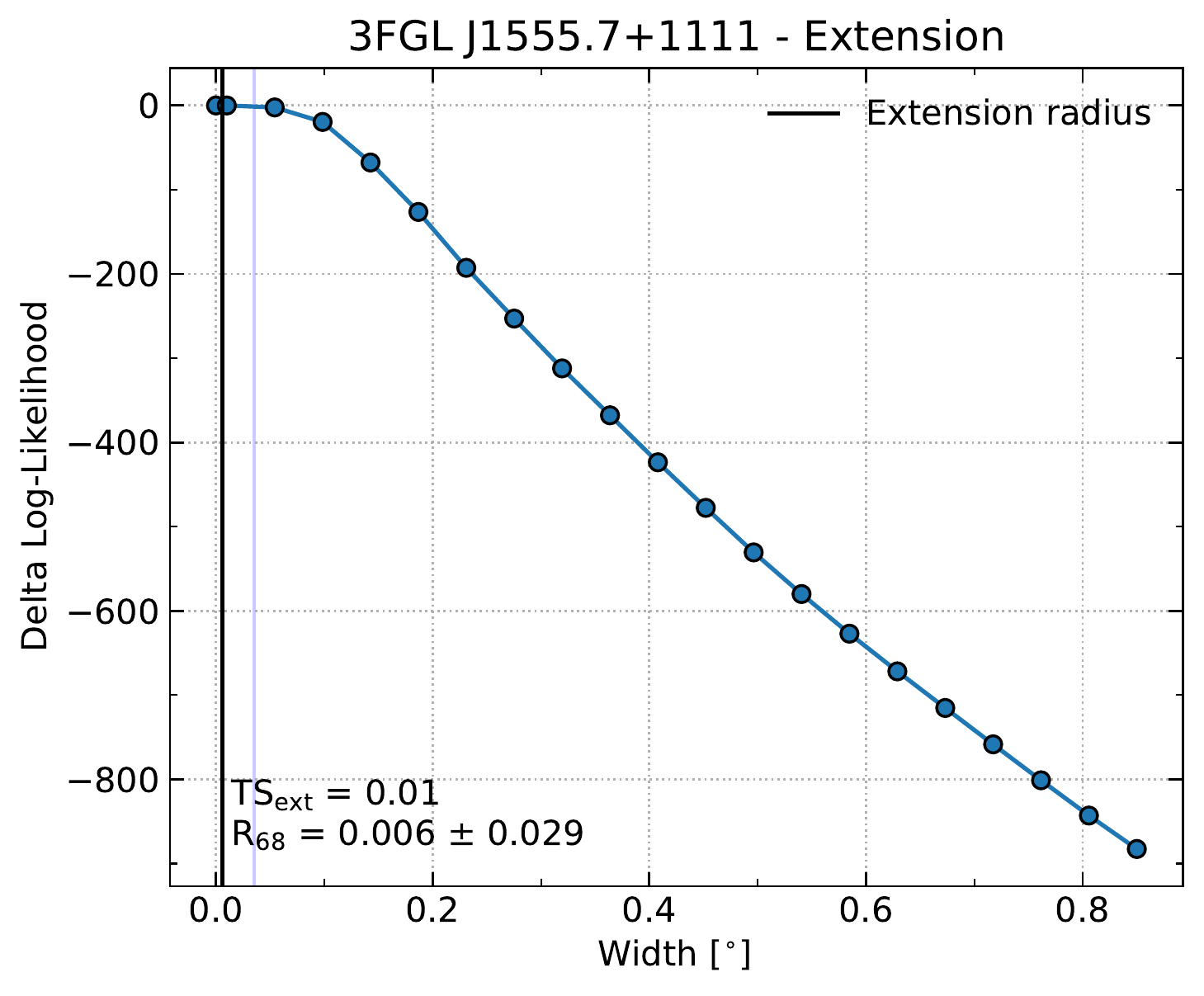}
    \caption{The extension of PG 1553+113 is compatible with zero, meaning that this source is point-like.}
    \label{fig:extension}
\end{figure}

In Figures \ref{fig:SED}, \ref{fig:LC}, and \ref{fig:eLC} we show the $\gamma$-ray spectrum and light curves for PG 1553+113 built with the dataset described in \S \ref{sec:installation}. Upper-limits are displayed whenever an energy or time bin has TS $\leq 9$. For the extension, since PG 1553+113 is a point-like source, the peak in the delta log-likelihood shown in Figure \ref{fig:extension} is centered at zero.

All of these results are saved in files named ``SOURCENAME\_task.fits'' and ``SOURCENAME\_task.npy'', where ``task'' here can be ``loc'', ``lightcurve'', ``SED'', or ``extension''. These files can easily be accessed with \texttt{numpy} \citep{harris2020numpy}, \texttt{astropy} \citep{2013astropy1,2018astropy2} and \texttt{TopCat} \citep{2005topcat}. Furthermore, \texttt{easyFermi} automatically plots the results as \texttt{.pdf} or \texttt{.png} figures labeled as, e.g. ``Quickplot\_task.pdf''.

\subsection{Goodness of fit}

The detection, flux determination and spectral modeling of \textit{Fermi}-LAT sources with \texttt{easyFermi} is accomplished by means of a binned likelihood optimization technique \citep{abdo2009fermi_likelihood}, using \texttt{MINUIT} as the minimizer. The multi-dimensional minimization of parameters, however, is not an easy task and is susceptible to fail. To be sure that the fit properly converged, the user can follow the information in the \textit{Log} box in the graphical interface (see Figure \ref{fig:easyFermi}), where the following information is displayed depending on the success of the fit:

\begin{itemize}
    \item Fit quality: 3. Excellent fit. Full accurate covariance matrix.
    \item Fit quality: 2. Reasonable fit. Full matrix, but forced positive-definite (i.e. not accurate).
    \item Fit quality: 1. Poor fit. Diagonal approximation only, not accurate.
    \item Fit quality: 0. Bad fit. Error matrix not calculated.
\end{itemize}

For any fit quality value other than 3, we recommend the user to rerun the analysis changing the configurations in the panel \textit{Free source radius}, in the ``Advanced configurations'' box. In this panel, the user can select the radius (from the ROI center) defining the circular region in which the parameters of the $\gamma$-ray sources are left free to vary, can fix the spectral shape of the target or of all sources in the ROI, and can freeze the Galactic and Isotropic diffuse emission models. For a step-by-step guide on the quality/goodness of fit, the user is invited to follow the tutorial on YouTube\footnote{\url{https://www.youtube.com/channel/UCeLCfEoWasUKky6CPNN_opQ}}.

\section{Validation, performance, and maintenance}

The results obtained with \texttt{easyFermi} are exactly the same as those obtained with \texttt{Fermipy}, since \texttt{easyFermi} is actually running  \texttt{Fermipy} in the background. We tested this by performing several analyses (e.g., the light curves, SEDs and TS maps for ROIs centered on PG1551+113, 3C 273, 3C279, Omega Centauri, and the extended lobes of Centaurus A) with \texttt{easyFermi} and comparing the results with pure \texttt{Fermipy} analyses. In terms of performance, an analysis done with \texttt{easyFermi} takes basically as much CPU time and RAM as an analysis done directly with \texttt{Fermipy} on the terminal, however, \texttt{easyFermi} is slightly faster and uses less RAM than an analysis performed with \texttt{Fermipy} in web-based interactive computing platforms. To test if \texttt{easyFermi} is properly working, we recommend the user to follow the tutorial on GitHub/YouTube\footnote{\url{https://github.com/ranieremenezes/easyFermi}}. Furthermore, the user can check if both \texttt{Fermipy} and \texttt{easyFermi} are working smoothly by simply typing

\texttt{\$ python}

\texttt{>> from fermipy.gtanalysis import GTAnalysis}

\texttt{>> import easyFermi}

If both modules are successfully imported, the installation is fine.

The maintenance of \texttt{easyFermi} will be done such that the users can ask for the repair of possible issues via GitHub and in a way that we guarantee its compatibility with the upcoming releases of the \texttt{Fermitools} and \texttt{Fermipy}.

\section{Conclusions}
\label{sec:conclusions}

The \texttt{easyFermi} graphical interface is a user-friendly tool that allows astronomers from all niches to use \textit{Fermi}-LAT $\gamma$-ray data. This tool is especially indicated for those scientists just starting in the field of high-energy astrophysics, and can be used for several goals like building a light curve, computing a $\gamma$-ray spectrum, or looking for extended $\gamma$-ray emission. Furthermore, \texttt{easyFermi} is meant to be simple. For more complex types of analyses, we refer the user to \texttt{Fermipy} and \texttt{Fermitools}. 

The tutorials and source code for \texttt{easyFermi} can be found online\footnote{\url{https://github.com/ranieremenezes/easyFermi}} and allow the user to learn how to use \textit{Fermi}-LAT data in just a few minutes. Bug reports and proposals for new functionality should be made through the GitHub issue tracker. Although minimum maintenance is required, we aim to keep \texttt{easyFermi} updated and in synergy with \texttt{Fermipy}, hence new versions of the tool can be released in the future. We also plan to make a similar graphical interface for the Cherenkov Telescope Array \citep{bernlohr2013_CTA} and possibly for other Imaging Atmospheric Cherenkov Telescopes in the near future.

\section*{Acknowledgements}

I would like to thank the anonymous referees for their suggestions and comments, as well as Alessandra Azzollini, Clodomir Vianna, Douglas Carlos, Fabio Cafardo, Kaori Nakashima, Lucas Costa Campos, Lucas Siconato, Ra\'i Menezes, Rodrigo Lang, and Romana Grossova for installing, testing, and helping me with the development of \texttt{easyFermi}. Part of this project was also supported by the European Research Council for the ERC Starting grant MessMapp, under contract no. 949555.

\section*{Data Availability}

All data used to exemplify the usage of \texttt{easyFermi} in this work can be found online in the \textit{Fermi}-LAT data server\footnote{\url{https://fermi.gsfc.nasa.gov/cgi-bin/ssc/LAT/LATDataQuery.cgi}}. 



\begin{thebibliography}{}

\bibitem{abdo2009fermi_likelihood} Abdo, Aous A., Ackermann, M., Ajello, M. et al. 2009, ApJS, 183, 46.

\bibitem{abdollahi2020_4FGL} Abdollahi, S., Acero, F., Ackermann, M. et al. 2020, Fermi large area telescope fourth source catalog, ApJS, 247, 33.

\bibitem{acero2015_3FGL} Acero, F., Ackermann, M., Ajello, M. et al. 2015, Fermi large area telescope third source catalog, ApJS, 218 (2), 23.

\bibitem{2013astropy1} Astropy Collaboration, Robitaille, Thomas P., Tollerud, Erik J., Greenfield, P. et al. 2013, Astropy: A community Python package for astronomy, Astronomy \& Astrophysics, 558, A33.

\bibitem{2018astropy2} Astropy Collaboration, Price-Whelan, A.M., Sip{\H{o}}cz, B.M., G{\"u}nther, H.~M. et al. 2018, The Astropy Project: Building an Open-science Project and Status of the v2.0 Core Package, The Astronomical Journal, 156 (3), 123.

\bibitem{atwood2009large} Atwood, W.B., Abdo, Aous A., Ackermann, M. et al. 2009, The large area telescope on the Fermi gamma-ray space telescope mission, ApJ, 697 (2), 1071.

\bibitem{bernlohr2013_CTA} Bernl{\"o}hr, K., Barnacka, A., Becherini, Y. et al. 2013, Monte Carlo design studies for the Cherenkov telescope array, Astroparticle Physics, 43, 171.

\bibitem{harris2020numpy} Harris, Charles R., Millman, K. Jarrod, van der Walt, St{\'{e}}fan J. et al. 2020, Array programming with NumPy, Nature, 585, 7825.

\bibitem{massaro2004log} Massaro, E., Perri, M., Giommi, P., Nesci, R. 2004, Log-parabolic spectra and particle acceleration in the BL Lac object Mkn 421: Spectral analysis of the complete BeppoSAX wide band X-ray data set, Astronomy \& Astrophysics, 413 (2), 489.

\bibitem{2005topcat} Taylor, M.B. 2005, TOPCAT \& STIL: Starlink Table/VOTable Processing Software, Astronomical Data Analysis Software and Systems XIV, Astronomical Society of the Pacific Conference Series, 347, 29.

\bibitem{wood2017fermipy} Wood, M., Caputo, R., Charles, E. et al. 2017, Fermipy: An open-source Python package for analysis of Fermi-LAT Data, 35th International Cosmic Ray Conference (ICRC2017), 301, 824.


\end{thebibliography}

\end{document}